# Coulomb-coupled quantum-dot thermal transistors

Yanchao Zhang, [1] Zhimin Yang, [1] Xin Zhang, [1] Bihong Lin, [2] Guoxing Lin, [1] and Jincan Chen[1,a]

[1]*Department of Physics, Xiamen University, Xiamen 361005, People's Republic of China*

[2]*College of Information Science and Engineering, Huaqiao University, Xiamen 361021, People's Republic of China*

**Abstract**

A quantum-dot thermal transistor consisting of three Coulomb-coupled quantum dots coupled to respective electronic reservoirs by tunnel contacts is established. The heat flows through the collector and emitter can be controlled by the temperature of the base. It is found that a small change in the base heat flow can induce a large heat flow change in the collector and emitter. The huge amplification factor can be obtained by optimizing the Coulomb interaction between the collector and the emitter or by decreasing the energy-dependent tunneling rate at the base. The proposed quantum-dot thermal transistor may open up potential applications in low-temperature solid-state thermal circuits at the nanoscale.

## I. INTRODUCTION

Controlling heat flow at the nanoscale has attracted significant attention because of its fundamental and potential applications.[1-3] The thermal diode effect and negative differential thermal resistance (NDTR) are two most important features for building the basic components of functional thermal devices, which are the key tools for the implementation of solid-state thermal circuits.[3,4] The first model of a thermal rectifier/diode was proposed by controlling the heat conduction in one dimensional nonlinear lattice.[5] Based on different microscopic mechanisms, a very significant rectifying effect was exhibited and the concept of NDTR was also proposed in the subsequent works.[6,7] In recent years, the thermal diode effect and NDTR have been extensively studied in the different systems including quantum-dot systems,[8-11] metal-dielectric interfaces,[12] metal or superconductor systems,[13-16] quantum Hall conductors,[17] and spin quantum systems.[18] One of the particularly interesting tasks is to further build and implement a thermal transistor, which is analogous to an electronic transistor and can control the heat flows at the collector and emitter by small changes in the temperature or the heat flow at the base. Since Li *et al.* put forward the first theoretical proposal for a thermal transistor,[7,19] several proposals have been given to design other types of thermal transistors, such as superconductor–normal-metal thermal transistors,[20] near-field thermal transistors,[21] far-field thermal transistors,[22-24] and quantum thermal transistors.[25] Moreover, new concepts for thermal devices such as thermal logical gates[26] and thermal memories[27-29] have also been proposed and demonstrated.

---

[a] Electronic mail: jcchen@xmu.edu.cn

In recent years, the electron and heat transport properties of Coulomb-coupled quantum-dot system have been investigated in detail in the thermoelectric generators[30,31] and refrigerators.[32] Moreover, recent experiments have shown that many new applications for Coulomb-coupled quantum-dot system including rectification,[33-35] logical stochastic resonance,[36] and thermal gating[37] can be realized by the voltage fluctuation or thermal fluctuation to control and manage the charge current. Ruokola *et al.* introduced a single-electron thermal diode consisting of two quantum dots in the Coulomb blockade regime. A remarkable feature of the Coulomb-coupled quantum-dot system is that the electron transport through the system is forbidden but the capacitive coupling between the two dots allows electronic fluctuations to transmit heat between the reservoirs.[9] Based on the Coulomb-coupled quantum dots, a thermal management device was recently constructed. A significant advantage of such a device is that it can not only implement thermal diodes separately in two different paths but also perform more thermal management operations, such as heat flow swap, thermal switch, and heat path selector.[11]

In this paper, we propose a thermal transistor consisting of three Coulomb-coupled quantum dots connected to respective electronic reservoirs. In our system, the electron transport between the quantum dots is forbidden and the heat transport by phonons is effectively suppressed at sub-Kelvin temperature.[9] But the heat transport between the reservoirs is allowed by electrons tunneling into and out of the dots and exchanging energy through Coulomb interaction. Thus, the present model is, in principle, a true thermal transistor. The concrete contents are organized as follows. In Sec. II, the model and basic principles of a Coulomb-coupled quantum-dot system are briefly described. In Sec. III, the NDTR and the thermal transistor effect are investigated. The influence of the energy-selective tunneling and the Coulomb interaction on the amplification factor is discussed in Sec. IV. Finally, the main results are summarized.

## II. THEORETICAL MODEL AND PRINCIPLE DESCRIPTION

The model of a quantum-dot thermal transistor (QDTT) is illustrated in Fig. 1. The QDTT is analogous to an electronic transistor consisting of three terminals: the base ($B$), the collector ($C$), and the emitter ($E$). Each terminal $a$ ($a = B, C, E$) consists of a electronic reservoir with temperature $T_a$ and a quantum dot with the lowest single-particle energy level $\varepsilon_a$ and the quantum dot is connected to the electronic reservoir through a tunnel barrier. Three quantum dots are capacitively coupled to each other and interact only through the long-range Coulomb force, so that there is no electron transport between the quantum dots. However, the Coulomb interaction $U_{ab}$ ($a, b = B, C, E$  $a \neq b$) between the quantum dots $a$ and $b$ allows through electron tunneling into and out of quantum dots $a$ and $b$ to transmit the heat between the electronic reservoirs $a$ and $b$ with a temperature difference. When a constant temperature bias is applied between the emitter and the collector, the heat flow between the emitter and the collector is generated by electrons tunneling into and out of the dots and exchanging energy through Coulomb interaction and can be fine-tuned by the temperature that is applied to the base. In principle, a good thermal transistor effect can be achieved by increasing the heat flow between the emitter and the collector or suppressing the base heat flow.

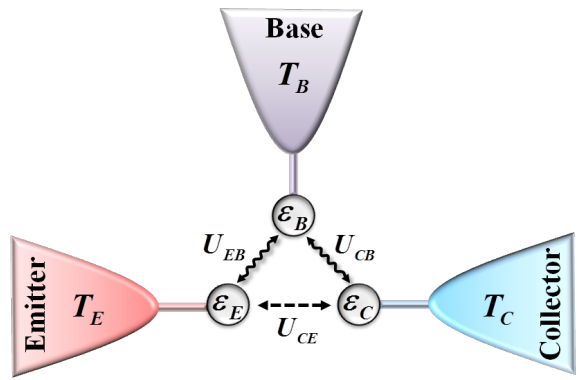

FIG. 1. The schematic diagram of a quantum-dot thermal transistor. Three Coulomb-coupled quantum dots are connected to respective electronic reservoirs. The Coulomb interaction $U_{CE}$ between the collector and the emitter is adjustable.

The Hamiltonian of the QDTT is described by

$$H = H_D + H_R + H_T, \qquad (1)$$

where

$$H_D = \sum_{a=B,C,E} \varepsilon_a d_a^\dagger d_a + \sum_{a,b=B,C,E \; a\neq b} U_{ab} d_a^\dagger d_a d_b^\dagger d_b \qquad (2)$$

is the Hamiltonian of Coulomb-coupled quantum dots, $(d^\dagger, d)$ denote the creation/annihilation operators of quantum dots,

$$H_R = \sum_k \sum_{a=B,C,E} \varepsilon_{ak} c_{ak}^\dagger c_{ak} \qquad (3)$$

is the Hamiltonian of reservoirs, $\varepsilon_{ak}$ is the energy of the noninteracting reservoir electrons with continuous wave number $k$, $(c^\dagger, c)$ denote the creation/annihilation operators of heat reservoirs,

$$H_T = \sum_k \sum_{a=B,C,E} \left( t_{ak} c_{ak}^\dagger d_a + t_{ak}^* d_a^\dagger c_{ak} \right) \qquad (4)$$

is the tunneling Hamiltonian between the quantum dots and the reservoirs, and $t_{ak}$ and its conjugate $t_{ak}^*$ denote the tunneling amplitudes.

The Coulomb-coupled quantum dot system is denoted by the charge configuration $|n_B, n_E, n_C\rangle$, where $n_a$ are the occupation number of quantum dots $a$. In the Coulomb blockade regime, each of these quantum dots can be occupied only by zero or one electron ($n_a = 0,1$). Thus, the dynamics of the quantum dot system is characterized by eight charge states labeled as $|1\rangle = |0,0,0\rangle$, $|2\rangle = |1,0,0\rangle$, $|3\rangle = |0,1,0\rangle$, $|4\rangle = |0,0,1\rangle$, $|5\rangle = |1,1,0\rangle$, $|6\rangle = |1,0,1\rangle$, $|7\rangle = |0,1,1\rangle$, and $|8\rangle = |1,1,1\rangle$. The occupation probabilities for eight charge states are given by the diagonal elements of the density matrix, $\boldsymbol{\rho} = (\rho_1, \rho_2, \rho_3, \rho_4, \rho_5, \rho_6, \rho_7, \rho_8)^T$. In the limit of weak tunneling coupling ($\hbar\gamma \ll k_B T$), the broadening of energy levels can be neglected and the transmission through tunnel barriers is well described by sequential tunneling rates. The off-diagonal density matrix elements do not contribute to steady state transport and can be neglected. Thus, the time evolution of occupation probabilities is given by a master equation. The matrix form can be written as $d\boldsymbol{\rho}/dt = \boldsymbol{M}\boldsymbol{\rho}$, where $\boldsymbol{M}$ denotes the matrix containing the transition rates and is given by Fermi's golden rule. The steady-state heat flows from electronic reservoir $a$ to quantum dot $a$ are given by the stationary solution of the master equation $\boldsymbol{M}\bar{\boldsymbol{\rho}} = 0$. Because the electronic reservoir of the emitter induces the transitions of charge states including $|1\rangle \leftrightarrow |3\rangle$, $|2\rangle \leftrightarrow |5\rangle$, $|4\rangle \leftrightarrow |7\rangle$ and $|6\rangle \leftrightarrow |8\rangle$, the heat flow of the emitter is given by

$$J_E = (\varepsilon_E - \mu_E)(\Gamma_{31}\bar{\rho}_1 - \Gamma_{13}\bar{\rho}_3) + (\varepsilon_E + U_{EB} - \mu_E)(\Gamma_{52}\bar{\rho}_2 - \Gamma_{25}\bar{\rho}_5) \\ + (\varepsilon_E + U_{CE} - \mu_E)(\Gamma_{74}\bar{\rho}_4 - \Gamma_{47}\bar{\rho}_7) + (\varepsilon_E + U_{EB} + U_{CE} - \mu_E)(\Gamma_{86}\bar{\rho}_6 - \Gamma_{68}\bar{\rho}_8).$$ (5)

The electronic reservoir of the collector triggers the transitions of charge states including $|1\rangle \leftrightarrow |4\rangle$, $|2\rangle \leftrightarrow |6\rangle$, $|3\rangle \leftrightarrow |7\rangle$ and $|5\rangle \leftrightarrow |8\rangle$, thus the heat flow of the collector is written as

$$J_C = (\varepsilon_C - \mu_C)(\Gamma_{41}\bar{\rho}_1 - \Gamma_{14}\bar{\rho}_4) + (\varepsilon_C + U_{CB} - \mu_C)(\Gamma_{62}\bar{\rho}_2 - \Gamma_{26}\bar{\rho}_6) \\ + (\varepsilon_C + U_{CE} - \mu_C)(\Gamma_{73}\bar{\rho}_3 - \Gamma_{37}\bar{\rho}_7) + (\varepsilon_C + U_{CB} + U_{CE} - \mu_C)(\Gamma_{85}\bar{\rho}_5 - \Gamma_{58}\bar{\rho}_8).$$ (6)

The electronic reservoir of the base drives the transitions of charge states including $|1\rangle \leftrightarrow |2\rangle$, $|3\rangle \leftrightarrow |5\rangle$, $|4\rangle \leftrightarrow |6\rangle$ and $|7\rangle \leftrightarrow |8\rangle$, and the heat flow of the base is expressed as

$$J_B = (\varepsilon_B - \mu_B)(\Gamma_{21}\bar{\rho}_1 - \Gamma_{12}\bar{\rho}_2) + (\varepsilon_B + U_{CB} - \mu_B)(\Gamma_{53}\bar{\rho}_3 - \Gamma_{35}\bar{\rho}_5) \\ + (\varepsilon_B + U_{EB} - \mu_B)(\Gamma_{64}\bar{\rho}_4 - \Gamma_{46}\bar{\rho}_6) + (\varepsilon_B + U_{CB} + U_{EB} - \mu_B)(\Gamma_{87}\bar{\rho}_7 - \Gamma_{78}\bar{\rho}_8).$$ (7)

In Eqs. (5)-(7), $\Gamma_{ji} = \gamma_a f_a(E_{ji})$ ($i,j = 1,2,\cdots,8$ and $i < j$) is the transition rate from the charge state $|i\rangle$ to $|j\rangle$ with an electron from the electronic reservoir $a$ into the quantum dot $a$, and $\Gamma_{ij} = \gamma_a(1 - f_a(E_{ji}))$ when an electron leaves the quantum dot $a$ into the electronic reservoir $a$, where $f_a(x) = \{\exp[(x - \mu_a)/(k_B T_a)] + 1\}^{-1}$ is the Fermi-Dirac distribution function with the chemical potential $\mu_a$ and temperature $T_a$, $E_{ji} = E_j - E_i$, $E_i$ is the energy of the charge state $i$, $k_B$ is the Boltzmann constant, and $\gamma_a$ is the energy-dependent tunneling rates between the quantum dots and the respective electronic reservoirs. In the present model, the electron transport between the quantum dots is forbidden. Through the etched trenches of about 150nm, the particle exchange between the quantum dot subsystems can be effectively prevented in the experiment.[33, 36] Hence, there is no Joule heating. Meanwhile, the heat dissipation by phonons is effectively suppressed at sub-Kelvin temperature.[9] Thus, the heat flows fulfill $J_C + J_E + J_B = 0$, which complies with the energy conservation.

### III. NDTR AND THERMAL TRANSISTOR EFFECT

Let us now turn to the discussion of a thermal transistor, where the small change in the base heat flow or the base temperature can control the heat flows through the emitter and collector. This thermal transistor effect is characterized by a amplification factor $\alpha_{E/C}$, which is defined as the change of the emitter heat flow or the collector heat flow divided by the variation of the heat flow applied at the base. In the present QDTT model shown in Fig. 1, the temperatures of the collector and emitter are fixed at $T_C$ and $T_E$ ($T_C < T_E$), respectively. The base at temperature $T_B$ ($T_C < T_B < T_E$) can be used to control the heat flows $J_E$ and $J_C$ with the help of the base heat flow $J_B$. Thus, the amplification factor $\alpha_{E/C}$ is defined as

$$\alpha_{E/C} = \frac{\partial J_{E/C}}{\partial J_B}. \tag{8}$$

When $|\alpha_{E/C}| > 1$, a thermal transistor effect will be observed. This means that the thermal transistor is able to amplify a small thermal signal. In other words, a small change of the heat flow though the base can yield a large change in the heat flows through the emitter and collector.

The differential thermal resistances (DTRs) of the emitter and collector are defined as

$$R_E = -\left(\frac{\partial J_E}{\partial T_B}\right)^{-1}_{T_E} \tag{9}$$

and

$$R_C = \left(\frac{\partial(-J_C)}{\partial T_B}\right)^{-1}_{T_C}. \tag{10}$$

By using Eqs. (9) and (10), the amplification factor in Eq. (8) can be rewritten as

$$\alpha_{E/C} = -\frac{R_{C/E}}{R_C + R_E}. \tag{11}$$

The thermal transistor effect, i.e., $|\alpha_{E/C}|>1$, implies that there exists a NDTR. The DTRs of the emitter and collector as a function of the base temperature $T_B$ are plotted in Fig. 2 (a).

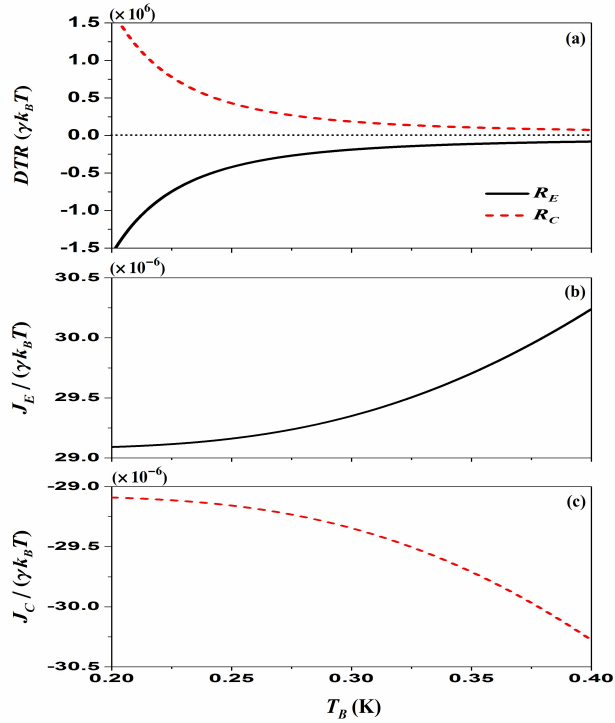

FIG. 2. (a) The differential thermal resistances of the emitter $R_E$ and the collector $R_C$ as functions of the base temperature $T_B$. (b) The emitter heat flow $J_E$ and (c) the collector heat flow $J_C$ as functions of the base temperature $T_B$. The parameters $\gamma_E = \gamma_C \equiv \gamma$, $\gamma_B/\gamma = 0.001$, $U_{EB} = U_{CB} = U_{CE} = 3k_BT$, $\mu_\alpha = 0$, $\varepsilon_E = 5k_BT$, $\varepsilon_B = 6k_BT$, $\varepsilon_C = -9.5k_BT$, $T_E = 400\text{mK}$, and $T_C \equiv T = 200\text{mK}$ are chosen.

It is shown that a NDTR is observed at the emitter. This then makes it possible that, over a wide regime of parameters, not only the emitter heat flow $J_E$ but also the collector heat flow $J_C$ may increase when the base temperature $T_B$ increase, as shown in Figs. 2 (b) and 2 (c). This is because that in the configuration of the present system, the available thermal energy from the collect reservoir at the low temperature $T_C$ is much smaller than the required energy $\mu_C - \varepsilon_C$. But, the probability of the electron tunneling into the collect reservoir is effectively controlled by the base temperature $T_B$ because the Coulomb interaction energy $U_{CB}$ can compensate for the energy required by the collect electron. When the base temperature $T_B$ is increased, the probability of the base electron tunneling increases. This can effectively energize the electron tunneling into the collect reservoir and thus efficiently transfer energy from emitter to collect through the Coulomb interaction energy $U_{CE}$. This asymmetric Coulomb blockade configuration is the origin of the NDTR.[9,11]

The thermal transistor effect is shown in Fig. 3. In Fig.3 (a), the base heat flow $J_B$ versus the base temperature $T_B$ is plotted. It is shown that the base heat flow $J_B$ is significantly smaller than the emitter heat flow $J_E$ and the collector heat flow $J_C$, and a small change of the heat flow $J_B$ corresponds to a large change of the heat flows $J_E$ and $J_C$. This leads to a noticeable amplification effect for the base heat flow $J_B$. The amplification ability of the thermal transistor is described by the amplification factors $\alpha_E$ and $\alpha_C$, as shown in Fig. 3 (b). It is found that the amplification factors $\alpha_E$ and $\alpha_C$ diverge and lead to an infinite amplification factor at $T_B^m \approx 285\text{mK}$, which is due to the fact that the base heat flow $J_B$ has a minimum at this point. In other words, the change rate of $J_B$ at this point is zero, as shown in Fig. 3 (a). In the range of $T_B > T_B^m$, the amplification factors $\alpha_E$ and $\alpha_C$ gradually decrease with the increase of the base temperature $T_B$. In particular, the base heat flow $J_B = 0$ at $T_B \approx 333\text{mK}$. This means that there is no heat flow though the base, but the amplification effect will still occurs at this point since $\alpha_E \approx 31$ and $\alpha_C \approx -32$. So within the range of these parameters, the system shows a good magnification effect.

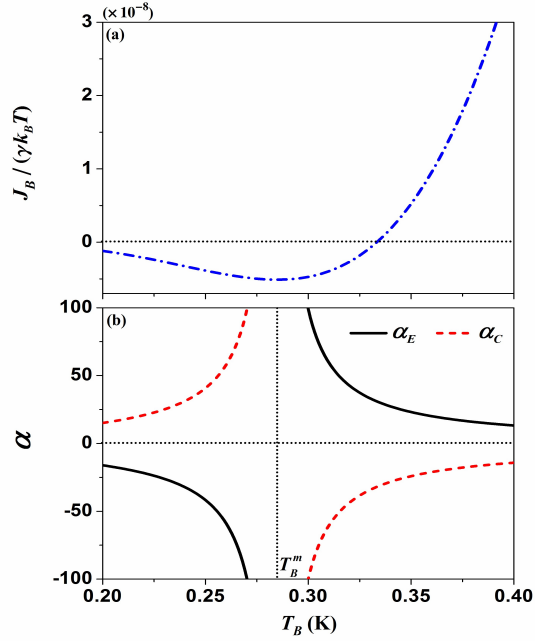

FIG. 3. (a) The base heat flow $J_B$ as a function of the base temperature $T_B$. (b) The amplification factors $\alpha_E$ and $\alpha_C$ as functions of the base temperature $T_B$. The values of other parameters are the same as those used in Fig. 2.

## IV. OPTIMIZATION OF AMPLIFICATION FACTORS

In such a device, cyclic transition enables the quantized packets of energy $U_{CE}$ to transfer from the emitter into the collector. However, there is a simultaneous transfer of the energy between the base and the emitter or collector. These are detrimental for obtaining thermal transistor effect. Therefore, in order to obtain good thermal transistor effect, two feasible methods can be implemented: (i) by reducing the base energy-dependent tunneling rate $\gamma_B$ or (ii) by adjusting the Coulomb interaction energy $U_{CE}$.

(i) Reducing the base energy-dependent tunneling rate $\gamma_B$ can effectively suppress energy transport at the base. This results in a decrease in the base heat flow, which can increase the amplification factor. Fig. 4 shows the amplification factors $\alpha_E$ and $\alpha_C$ as functions of the base tunneling rate $\gamma_B/\gamma$. It is found that the decrease of the base energy-dependent tunneling rate $\gamma_B$ can significantly increase the amplification factors $\alpha_E$ and $\alpha_C$. Actually energy-dependent tunneling rates typically occur quite naturally in top-gate-defined quantum dot structures.[38] The experimental demonstration of the direct control of energy-dependent tunneling rates is well suited to this purpose.[31]

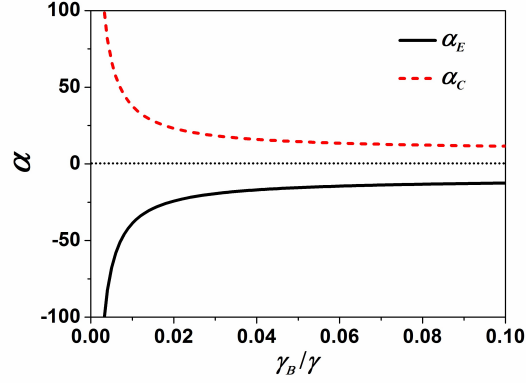

FIG. 4. The amplification factors $\alpha_E$ and $\alpha_C$ as functions of the base tunneling rate $\gamma_B/\gamma$ for $T_B = 280\text{mK}$. The values of other parameters are the same as those used in Fig. 2.

(ii) Adjusting the Coulomb interaction $U_{CE}$ between the collector and the emitter can directly control the quantized packets of energy $U_{CE}$ to transfer from the emitter into the collector. Adjusting the Coulomb interaction $U_{CE}$ can be done experimentally by electrostatic bridging of two subsystems.[39,40] The DTRs of the emitter $R_E$ and the collector $R_C$ as functions of $U_{CE}$ are shown in Fig. 5 (a). It is clear shown that the DTRs of the emitter $R_E$ and the collector $R_C$ flip each other at the point of $U_{CE}^0 = 8.1 k_B T$. At this point, the amplification factors $\alpha_E$ and $\alpha_C$ are zero at the same time, as shown in Fig. 5 (b), which shows the curves of amplification factors $\alpha_E$ and $\alpha_C$ versus the Coulomb interaction $U_{CE}$. It is found that $U_{CE}$ has an optimized interval, which can give a significantly greater amplification factor. But, the amplification factors $\alpha_E$ and $\alpha_C$ gradually decrease with the increase of $U_{CE}$ and eventually lead to the loss of amplification effect. When the Coulomb interaction $U_{CE} = 0$, there will be no heat transfer from the emitter to the collector. In such a case, the system will lose the thermal transistor effect.

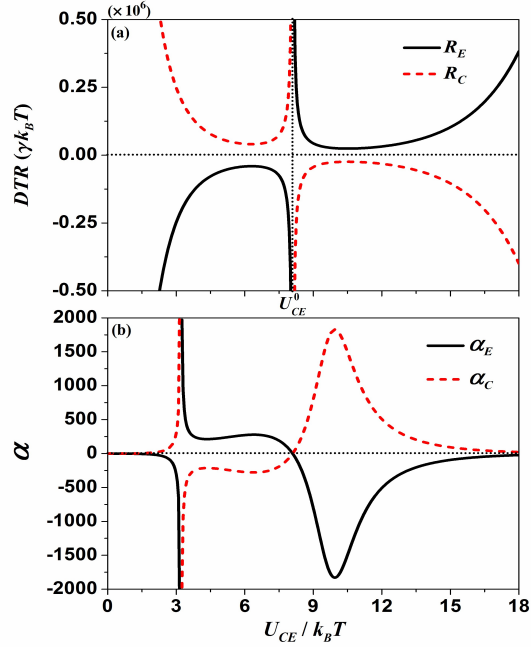

FIG. 5. (a) The differential thermal resistances of the emitter $R_E$ and the collector $R_C$ as functions of $U_{CE}$ for $T_B = 280\text{mK}$. (b) The amplification factors $\alpha_E$ and $\alpha_C$ as functions of $U_{CE}$ for $T_B = 280\text{mK}$. The values of other parameters are the same as those used in Fig. 2.

As an example, the amplification factors $\alpha_E$ and $\alpha_C$ as functions of the base temperature $T_B$ for $U_{CE} = 10k_BT$ are illustrated in Fig. 6. It is found that in the entire interval of the base temperature $T_B$, the amplification effect of the heat transistor is very significant. Even though the proposed model is based on the same general idea in Ref. [25], the underlying microscopic mechanism in the present model is different from that in Ref. [25]. The amplification factors $\alpha_{E/C}$ are larger than those in the model of the near-field thermal transistor[21] and quantum thermal transistor.[25] Recently, the Coulomb-coupled quantum dots system has been experimentally realized and extensively studied.[31,33,36,37] Therefore, our work will be sufficiently large meaning to encourage the experimental work in the near further.

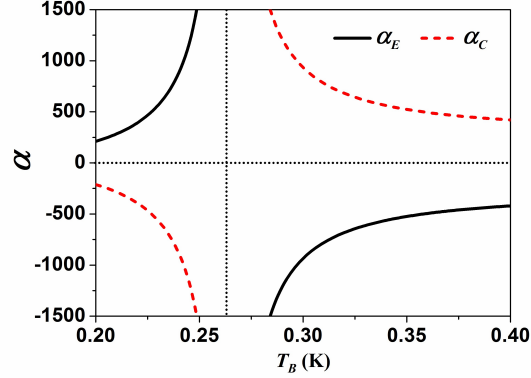

FIG. 6. The amplification factors $\alpha_E$ and $\alpha_C$ as functions of the base temperature $T_B$ for $\gamma_B/\gamma = 0.001$ and $U_{CE} = 10 k_B T$. The values of other parameters are the same as those used in Fig. 2.

## V. CONCLUSIONS

The performance of a thermal transistor consisting of three Coulomb-coupled quantum dots has been theoretically analyzed. It is found that the phenomenon of the NDTR constitutes the main ingredient for the operation of a thermal transistor. A small change in the heat flow through the base can control a large heat flow change in the collector and emitter. Such a thermal transistor is able to amplify a small heat signal that is injected into the base. The thermal transistor effect can be significantly improved by reducing the base energy-dependent tunneling rate or optimizing the Coulomb interaction between the collector and the emitter. Extremely high amplification factors can be obtained in a wide range of the base temperature. These results will stimulate interest for the quantum-dot thermal transistor and open up potential applications in low-temperature solid-state thermal circuits at the nanoscale.

## ACKNOWLEDGMENT

This work was supported by the National Natural Science Foundation (No. 11675132), People's Republic of China